# Title: Knowledge-guided Contextual Gene Set Analysis Using Large Language Models

**Authors:** Zhizheng Wang[1,‡], Chi-Ping Day[2,‡], Chih-Hsuan Wei[1], Qiao Jin[1], Robert Leaman[1], Yifan Yang[1], Shubo Tian[1], Aodong Qiu[3], Yin Fang[1], Qingqing Zhu[1], Xinghua Lu[3], Zhiyong Lu[1,*]

**Affiliations:**

[1] Division of Intramural Research (DIR), National Library of Medicine (NLM), National Institutes of Health (NIH); Bethesda, MD 20894, USA

[2] Cancer Data Science Laboratory, Center for Cancer Research, National Cancer Institute (NCI), National Institutes of Health (NIH); Bethesda, MD 20894, USA

[3] Department of Biomedical Informatics, School of Medicine, University of Pittsburgh; Pittsburgh, PA 15206, USA

‡ These authors contributed equally to this work

* Correspondence: zhiyong.lu@nih.gov

**Abstract:** Gene set analysis (GSA) is a foundational approach for interpreting genomic data of diseases by linking genes to biological processes. However, conventional GSA methods overlook clinical context of the analyses, often generating long lists of enriched pathways with redundant, nonspecific, or irrelevant results. Interpreting these requires extensive, ad-hoc manual effort, reducing both reliability and reproducibility. To address this limitation, we introduce cGSA, a novel AI-driven framework that enhances GSA by incorporating context-aware pathway prioritization. cGSA integrates gene cluster detection, enrichment analysis, and large language models to identify pathways that are not only statistically significant but also biologically meaningful. Benchmarking on 102 manually curated gene sets across 19 diseases and ten disease-related biological mechanisms shows that cGSA outperforms baseline methods by over 30%, with expert validation confirming its increased precision and interpretability. Two independent case studies in melanoma and breast cancer further demonstrate its potential to uncover context-specific insights and support targeted hypothesis generation.

**Main Text:**

## INTRODUCTION

Genomic, epigenetic, and transcriptomic analyses greatly facilitate our understanding of biological processes and mechanisms in both normal organism development and disease[1,2,3,4,5]. These analyses generate data for the expression, regulation, or alteration of a series of genes, so the results often include gene sets. For instance, exome sequencing allows mutation calling in coding genes[6,7], RNA sequencing reads gene expression levels[8,9], and ATAC-seq (assay for transposase-accessible chromatin sequencing) identifies transcriptionally active genes[10,11].

Functional interpretation of such gene sets is central to genomic research, enabling scientists to formulate hypotheses regarding disease mechanisms, potential biomarkers, and therapeutic approaches[12,13,14]. Gene set analysis (GSA)[15,16] is a widely used approach for identifying biologically significant pathways within gene sets, offering critical insights into the molecular mechanisms and biological processes underlying genomic data. Such mechanistic insights are generally crucial for uncovering disease threptic targets. Traditional GSA methods[17,18,19] typically compare gene sets against predefined categories within manually curated databases, making them widely applicable to high-throughput omics datasets, including transcriptomics, proteomics, and single-cell sequencing, by identifying functions that are statistically significantly enriched. More recently, GSA methods leveraging large language models (LLMs)[20,21,22] have demonstrated promise by generating meaningful functional annotations. Through instruction-based learning and self-verification strategies, LLM-powered approaches provide coherent biological narratives for gene sets, further enhancing the utility of GSA in contemporary clinical studies.

Nevertheless, these GSA methods do not account for experimental contexts or study objectives associated with the diseases, often producing generic, fragmented, or redundant results. This issue is particularly pronounced in high-throughput omics studies, where the biological relevance of enriched pathways is highly sensitive to experimental conditions[23,24]. Lacking mechanisms for prioritizing contextually relevant results, researchers frequently encounter extensive lists of enriched pathways. However, many of these enriched pathways are unrelated to the experimental conditions and study objectives[25,26]. Consequently, results are often filtered in an ad-hoc manner, such as only selecting statistically top-ranked pathways, which risks overlooking contextually relevant pathways that are buried at the bottom of the enrichment list.

Furthermore, current GSA methods overlook gene interactions and regulatory relationships[27] within gene sets, resulting in central hub genes over-dominating enrichment results regardless of their actual mechanistic involvement. For instance, the tumor suppressor gene *TP53* is frequently identified as very significantly enriched across numerous cancer-related pathways[28,29,30], even when many of these pathways are not directly relevant to a specific cancer type. This bias not only inaccurately inflates the significance of hub genes but also limits the discovery of functionally relevant pathways that may be critical to the research objectives yet remain underexplored.

To overcome these limitations and enhance the interpretation of function related to specific research objectives, we present **Contextualized Gene Set Analysis** (cGSA), a novel LLM-based pipeline that integrates both domain knowledge and study context into an LLM reasoning process for gene set enrichment analysis. Specifically, cGSA features three key technical innovations:

**Context-aware GSA**: Incorporates contextual information to tailor results to specific study objectives and experimental conditions, significantly improving relevance and accuracy, thereby reducing background noise.

**Gene Network Clustering**: Clusters genes via protein-protein interaction (PPI) networks and community detection algorithms, enhancing pathway specificity and reducing the over-inflated enrichment of central hub genes.

**Interpretable Results via Instruction and Summarization**: Guides LLMs to generate concise, biologically meaningful, and contextually relevant pathways via pertinent summarization, minimizing potential hallucinations and over-generalizations.

Moreover, to fill the gap that current benchmarks used in existing GSA methods merely include co-expressed genes, we manually curated a new benchmark dataset of differentially expressed genes (DEGs) with explicit contextual information of multiple common diseases and cancers (**Data file S1, Evaluated DEGs**), which is central to enrichment analysis.

Finally, to assess the real-world utility of our method, we applied cGSA to two practical research studies: development of resistant to immune checkpoint blockade (ICB) in melanoma[31] and cell-cell communication within breast cancer tumor microenvironment[32]. Expert evaluation confirms that cGSA enhances functional interpretations by aligning with established findings or making connections with contextually relevant knowledge. We make cGSA accessible through a publicly available demo website and promote transparency and reproducibility by sharing our detailed GPT instructions.

## RESULTS

### Overview of the cGSA Workflow

cGSA is an LLM-based gene set analysis pipeline designed to identify context-specific functional pathways for DEGs (**Figure 1a**). Its workflow comprises four key steps (**Materials and Methods, Figure 1b**). First, cGSA identifies interacting genes within the input DEGs using the STRING database[33] and applies the EdMot algorithm[34] to partition them into non-overlapping clusters, capturing distinct co-abundance gene subsets. Second, it performs classical enrichment analysis on each gene cluster using the Enrichr API[18], applying a p-value threshold ($\alpha = 0.001$) to identify significantly enriched pathways. Third, an LLM such as GPT-4o[35] filters the predicted functional pathways by considering the user-provided experimental context where the DEGs were generated, allowing cGSA to remove contextually irrelevant results. Finally, the LLM identifies the most representative pathway for each cluster from all experimentally relevant candidates, guided by the user-provided overall study objectives. It summarizes the known biological functions of these pathways and assigns a confidence score to indicate their contextual relevance. By incorporating user-provided context in the filtering and summarization steps, cGSA improves the relevance of its pathway predictions and functional interpretations to the specific experimental conditions and research goals. The prompts used in LLMs are shown in **Data file S2**.

**cGSA Significantly Outperforms Current Methods**

We conducted a comprehensive performance comparison to evaluate the effectiveness of cGSA and its ability to overcome the limitations of traditional enrichment analysis and LLM-based methods. This evaluation is essential for determining whether cGSA reliably generates contextually relevant and biologically meaningful pathways while minimizing irrelevant outputs. For evaluation, we manually curated over 102 DEG sets (**Table 1**) spanning 19 cancer-related diseases (e.g., *melanoma*) and ten disease-related biological mechanisms (e.g., *tumor immune microenvironment in hepatocellular carcinoma*) from peer-reviewed publications. Each DEG set with the averaged gene number of 914, derived from the original studies, is known to be associated with well-defined contextual biological functions, serving as the ground truths in this study. These contextual annotations provide essential details for interpreting significant gene expression changes of DEGs under distinct biological or experimental conditions[36,37].

Specifically, we compared cGSA against four baselines, including three advanced LLMs (GPT-4o[35], GPT-4[38], and Llama 3.1-70B[39]) and a traditional gene set enrichment analysis method, g:Profiler[17]. For a fair comparison, the LLM-based GSA methods followed the prompting strategy proposed by Joachimiak *et al.*[40] for summarizing the plausible biological terms for a given gene set, while g:Profiler used all databases specified in its original publication. Importantly, none of the baseline methods is able to incorporate contextual information.

Directly applying LLMs to this task generates pathway names as free text, which may be semantically similar to the ground truth but not an exact word-by-word match. For example, LLMs generate the pathway name "glycine cleavage system" for *GO:0019464* whose GO term is "glycine decarboxylation via glycine cleavage system". We can see that they are not exactly matched but with high semantical similarity. To enable a more meaningful evaluation, we employed semantic similarity scores to measure the alignment between the predicted terms and the ground truth annotations. Model performance was assessed using two metrics (**Materials and Methods**): accuracy (ACC) – the proportion of functionally related pathways among all cGSA generated pathways – and the hit ratio (HIT) – the proportion of ground truths matched by generated predictions. Performance was summarized using the harmonic mean of ACC and HIT across different similarity score thresholds (0.5 - 0.9). These similarity scores were calculated as the cosine similarity between the embedding vectors of the ground truth annotations and the predicted functional pathways, as computed by MedCPT[41], a state-of-the-art biomedical semantic encoder. For instance, the pathway "focal adhesion" predicted by cGSA exhibits a high similarity score of 0.94 to its ground truth counterpart, "focal adhesion assembly," while displaying a substantially lower similarity score of 0.57 with another ground truth pathway, "Axon guidance."

At a similarity score threshold of 0.5, which best aligns with the human verification result (see details in the section below **High Consistency Between Automatic and Expert Annotation**), cGSA achieves a harmonic mean of 0.878, representing a 30% improvement over g:Profiler, 33% over Llama 3.1-70B, and 37% improvement over GPT-4o. Across similarity score thresholds ranging from 0.5 to 0.9, cGSA maintains an average harmonic mean of 44.3%, significantly outperforming the baseline methods (**Figure 2a**). These results highlight cGSA's ability to generate pathways that are more closely aligned with ground truths by generating fewer but more precise outputs.

On average, cGSA generated 14 pathways per input DEG set, compared to 683 pathways predicted by the g:Profiler (**Figure 2b**). Despite a substantial reduction, cGSA achieved a HIT value 21% higher than g:Profiler (**Data file S7**), indicating that it recalled more ground truth pathways despite its conservative output. As a result, cGSA achieved a significantly higher ACC than g:Profiler.

To align with common research practices, where researchers typically streamline the validation by prioritizing the most enriched pathways, we compared cGSA with the top K (K = 10, 30, 50) results from g:Profiler (**Figure 2c**). The performance of g:Profiler declined by approximately 11% at K = 50 and about 27% at K = 10. Moreover, its HIT values peaked (61.2% vs. 63.8%) only when the number of candidate pathways was expanded to 50. These findings suggest that many contextually relevant pathways for a given gene set are indeed ranked lower in enrichment analysis outputs, necessitating extensive manual review of the entire result list in order not to avoid missing key pathways. In contrast, cGSA effectively prioritizes contextually relevant pathways in a more concise manner, significantly reducing the manual review effort.

## High Consistency Between Automatic and Expert Annotation

To assess the validity of the automatic evaluation method based on similarity scores, we recruited two human experts to examine the functional associations between 1,469 cGSA-generated pathways and 1,671 ground truth pathways in the 102 manually annotated DEG sets (**Data file S3**). The experts demonstrated a high inter-annotator agreement of 90.0%. This evaluation also employed the ACC and HIT metrics (**Figure 3a**). Please note that manual annotations were limited to our cGSA generated pathways (instead of predictions by all other competing methods) due to its high cost.

According to the human validation, cGSA achieved an ACC of 0.828, a HIT of 0.873, and a harmonic mean of 0.850, closely aligning with the result of 0.878 obtained using a similarity score threshold of 0.5 (**Figure 2a**). To further validate the reliability of the 0.5 similarity threshold, we computed the MedCPT similarity score for all functionally related pathways annotated by the human experts (**Figure 3b**). This calculation yielded an average similarity of 0.543, further supporting the use of the 0.5 threshold in the previous automatic method comparison analysis.

Additionally, comparing the relevance scores of functionally related and unrelated pathways (**Figure 3c**) revealed that functionally related pathways had an average relevance score of 7.0, whereas unrelated pathways averaged 5.3. This finding suggests that a relevance score above 7.0 strongly supports contextual relevance. Furthermore, we examined the correlation between relevance scores and the number of genes recorded in the database for each pathway (**Figure 3d**). This analysis revealed that pathways with fewer recorded genes generally had lower relevance scores (below 5.5), while those with a larger number of recorded genes tended to have higher relevance scores (above 5.5). Based on these findings, we established two cutoff thresholds: 5.5 for moderate contextual relevance and 7.0 for high contextual relevance.

**cGSA is Robust with the Choice of Different LLMs and Contextual Terms**

To assess the contribution of each key component in cGSA, we conducted ablation experiments focusing on gene cluster detection, backend LLMs, and the sensitivity of cGSA to context length.

**Gene Cluster Detection**. To evaluate the role of gene cluster detection in mitigating the inflated enrichment caused by central hub genes, we removed this component and instead prompted the LLM to directly summarize context-related pathways based on the enrichment analysis of the entire input DEG set. The results (**Figure 4a**) show that gene cluster detection enhances cGSA's performance by 16.4%, primarily by identifying distinct gene subsets within the list of DEGs and improving ground truth coverage.

For example, in the analysis of DEGs collected from PMC11350455[42], cGSA without gene cluster detection produced four pathways: "Complement", "KRAS.50 UP.V1 UP (MSigDB)", "KRAS.KIDNEY UP.V1 UP (MSigDB)", and "KRAS.300 UP.V1 UP (MSigDB)". Despite differences in nomenclature, these pathways largely represent only two functional categories: complement activation and KRAS signaling. This redundancy was caused by the disproportionate influence of the central hub genes *SCG5* and *SCG3*, leading to functionally overlapping results. In contrast, cGSA with gene cluster detection identified three distinct and contextually relevant pathways: "mTORC1 signaling pathway," "KRAS signaling pathway," and "Oxidative Phosphorylation" (**Data file S3, Functional Category Annotation**). These results not only retained the prominent biological process of KRAS signaling but also captured a broader set of ground truth pathways, resulting in a higher HIT score and improved biological interpretability.

**LLM Backend**. We further evaluated the impact of different LLMs (GPT vs. Llama) on the pathway screening and pathway summarization steps. Compared to Llama 3.1-70B, cGSA configured with GPT models demonstrated superior performance in summarizing context-related pathways (**Figure 4b** and **4c**). This effect arises from differences in response behavior between GPT and Llama models when using the fully engineered prompt (**Data file S2**). The GPT models tend to prioritize summarizing a single prominent functional term. In contrast, the Llama models are more likely to concatenate multiple significant pathways rather than selecting the most representative one. For example, in a study related to melanoma (PMID30017245[43]), the Llama model generated the term "Regulation of Cell Migration and Angiogenesis," whereas the GPT model summarized it as "Regulation of MAPK Cascade". The latter exhibits greater semantic relevance to the ground truth, highlighting GPT's tendency to produce more contextually precise pathway interpretations.

Additionally, the harmonic means of ACC and HIT revealed no significant difference between cGSA using GPT-4o and GPT-4. Furthermore, cGSA performance is also consistent on the subset of input DEGs published in 2024 versus those in 2022-23. This result suggests that cGSA is not highly sensitive to the knowledge cutoff of different GPT-4 versions, as our first gene cluster detection step effectively breaks input DEGs into unknown gene subsets using the community detection method EdMot.

**The sensitivity of cGSA to different length of contextual terms**. To examine the impact of context length on pathway summarization, we tested cGSA using three different context lengths (**Figure 4d**). In addition to the original context extracted from published articles, with an average length of 6.92 words, we generated keyword (KW) sequences with an average length of 2.59 words and curated professional statements with an average length of 8.71 words based on the original contexts. For instance, in the study associated with PMC11350455[42] (**Data file S1**, **Evaluated DEGs**), the research objective linked to the DEGs was "pancreatic neuroendocrine tumors." Accordingly, the keyword was "pancreatic neuroendocrine," while the professional statement provided a more detailed description: "Clinically aggressive and nonaggressive proteome subtypes of pancreatic neuroendocrine tumors." The results, ranging from 0.871 to 0.880, indicate that cGSA remains robust across varying context lengths, including very short contexts consisting of only a few key terms.

Overall, these experiments confirm that each component of cGSA contributes to its performance and robustness. Gene cluster detection significantly enhances accuracy and coverage, while the use of curated target databases and GPT-based LLMs improves accuracy of enrichment analysis and pathway summarization. Furthermore, cGSA's robustness to variations in context length underscores its adaptability across diverse experimental settings.

**Two Real-World Case Studies using cGSA**

**cGSA assists analysis of adaptive gene expression in melanoma evolution to identify functions selected for therapeutic resistance.**

In a recent study, Hirsch *et al.*[31] identified the adaptively up- and down-regulated genes (AUGs and DUGs, respectively) in two specific subclone groups of a parental mouse melanoma B2905 cell line, one group exhibiting an invasive phenotype and the other a proliferative phenotype, (**Figure S1, step 1 to 4**). They further analyzed gene expression of the parental B2905 melanoma treated with an immune checkpoint blockade (ICB) therapy (**Figure S1, step 5**), suggesting that the AUGs (**Data file S1, Case-Study DEGs**) of the invasive subclone group were selected by therapy-induced immunity, confirming its function of resistance to ICB. However, whether adaptive genes of other subclones in the parental cell line are also subjected to selection, and what their functions would be, remains unclear. To validate the utility of cGSA in such studies, we applied the pipeline to analyze the AUGs and DUGs of the subclone groups with invasive and proliferative phenotype information (**Figure S2**), as well as DEGs between responding (R) and non-responding (NR) B2905 tumors receiving ICB treatment in a preclinical study (**Figure 5a**).

For comparison, we categorized the cGSA-generated pathways into nine functional groups and color-coded them, as shown in **Figure 5a**. The genes expressed at higher levels in responding tumors (R > NR) are enriched in pathways of category 1 (metabolism), 2 (transcriptional adaptation), 3 (cell migration), 5 (calcium signaling), and 7 (immune response). In contrast, the genes expressed at higher levels in non-responding tumors (NR > R) are mostly enriched in pathways of category 6 (cell cycle) with a total of 70 genes and higher relevance scores (7.5-8.5). Other enriched pathways have 13 genes in total, with lower relevance scores (6.0-7.5) (**Figure 5a**). The analysis of R > NR genes demonstrated that ICB induced an effective immune response in the responding tumors, as revealed by the category 7 pathways, and the immunity selected genes for the phenotypes aligned with the invasive subclone groups, whose AUGs were also specifically enriched in the pathways of category 1, 2, and 3[31] (**Figure S2, a**). In fact, the pathway enriched by the greatest number of genes in both cases are "Regulation of Transcription by RNA Polymerase II", with 33 genes (score= 7.5) in the former and 41 genes (score = 7.5) in the latter (**Figure 5a**). These results suggested that these functional categories drive resistance to ICB therapy. In contrast, the NR > R genes are dominantly enriched in pathways of cell cycle function. This is more aligned with the phenotype of the proliferative subclone group (**Figure S2, b**), whose adaptive upregulated genes were enriched in the pathways of cell growth function, including category 4 (mitochondrial function) and 7 (translation). Both cases have enrichment in mitochondrial function.

As the study carried out by Hirsch *et al*. focused on the specific genes with adaptive expression, the results assisted by cGSA further reveal the functions of the genes whose expression was under selection. We observed that the same categories of pathways, from our analysis of subclone groups and ICB-treated tumors have minimally overlapping genes. This demonstrates the capacity of cGSA in functional interpretation of the input gene sets, instead of simple statistical comparison between input and established database. Comprehensive outcomes of cGSA for this case study are detailed in **Data file S4, Case Study 1**.

**cGSA helps reveal pathways causally impacted by cell-cell communication within the breast cancer tumor microenvironment.**

Cell-cell communication (CCC) within the tumor microenvironment (TME) is crucial for regulating immune responses, promoting tumor progression, and influencing therapeutic targets by mediating interactions between immune, stromal, and tumor cells through receptor-ligand signaling, cytokines, and growth factors[44,45,46]. Immune cells, such as T cells and myeloid cells, engage in CCC to modulate anti-tumor immunity, while endothelial cells regulate angiogenesis through VEGF signaling, and cancer-associated fibroblasts contribute to TME remodeling[47,48]. Given the importance of CCC in shaping treatment responses, we evaluate the applicability of cGSA in analyzing CCC within the TME by applying the pipeline to DEGs identified from on-treatment versus pre-treatment comparisons across T cells, myeloid cells, and endothelial cells. These DEGs (**Data file S1, Case-Study DEGs**) were derived through single-cell sequencing data from 31 breast cancer patients receiving anti-PD1 therapy[32].

By considering the context in our study, cGSA identified TNF-α/NF-κB signaling in the T cells and endothelial cells, and G protein-coupled receptor (specifically, GPCR-cAMP) signaling in myeloid cells (**Figure 5b**). Importantly, the genes enriched in these pathways are linked to the inflammatory states of immune cells based on causal knowledge, suggesting their role in cell-cell interactions, and providing candidates for further study. For example, our result suggests an activation-suppression connection between *FKBP5* in the TNF-α signaling pathway (via NF-κB) in T cells and the suppression of *HCAR2* in the GPCR-cAMP signaling pathway in myeloid cells (**Figure 5b, upper panel**). This relationship aligns with the mechanism of anti-PD1 therapy, which activates inflammatory signaling via pathways such as TNF-α/NF-κB in T cells, leading to the upregulation of *FKBP5*[49,50]. Increased *FKBP5* expression, in turn, stabilizes and prolongs NF-κB activation[51,52,53], enhancing the release of pro-inflammatory cytokines (e.g., TNF-α, IL-1, and IL-6). Pro-inflammatory conditions, including TNF-α and IL-6 signaling, can create an environment where *HCAR2* activation is reduced, potentially contributing to a shift in myeloid cells toward a more inflammatory phenotype[54,55]. It also suggests a causal connection between *TRAF1* in T cells and *NFKBIZ* in endothelial cells, both participating in the TNF-α signaling via NF-κB (**Figure 5b, lower panel**). A plausible explanation for this finding is anti-PD1 treatment activating T cells, which upregulates *TRAF1* and modulates NF-κB signaling[56]. Activated T cells then secrete TNF-α, which binds *TNFR1* on endothelial cells, triggering NF-κB signaling and inducing NFKBIZ expression[57,58]. *NFKBIZ* enhances transcription of inflammatory genes like *IL-6*, *ICAM-1*, and *CXCL10*, influencing endothelial cells[59,60]. These cells recruit immune cells, sustaining T-cell activation and cytokine secretion, creating a positive feedback loop[61].

Our results demonstrate that cGSA provides more specific pathways relevant to both research objectives and cell types. Integrating cGSA with causal inference analyses can yield valuable insights on the mechanisms driving anti-PD1 therapy, generating useful hypothesis for experimental validation. Comprehensive outcomes of cGSA for this case study are detailed in **Data file S4, Case Study 2**.

## DISCUSSION

### Importance of Context in Gene Set Analysis

The contextual information of DEGs is crucial for biologically grounded interpretations and ensuring that analyses align with the overarching goals of the study[23,24]. However, conventional enrichment analysis tools always lack mechanisms to specify research objectives, often producing substantial lists of predicted pathways with limited relevance. As a result, users must manually identify the relevant pathways – a subjective and ad-hoc process which decreases the reproducibility of findings.

Large language models provide a promising approach for incorporating context into functional enrichment analysis, given their powerful text summarization capability built upon huge training data where have lot of text discussing gene associations with functions. To enhance both accuracy and contextual relevance, cGSA integrates LLMs with conventional gene set enrichment analysis through a summarization-based strategy as shown in **Figure 1b**. Rather than generating a functional term from scratch, cGSA refines standard enrichment results by selecting and summarizing contextually relevant pathways.

This approach addresses two major challenges effectively. First, it makes LLMs clearly understand the task, preventing overly cautious responses, such as "As an AI, I cannot perform enrichment analysis for gene sets". Second, it mitigates fragmented outputs that merely list multiple related pathways (e.g., "MAPK signaling, Ras signaling, JAK-STAT signaling, …") without synthesizing them into a coherent biological interpretation centered on the most representative biological process.

**Error Analysis**

While cGSA demonstrated superior performance, it missed 16.5% (275 out of 1671) of ground truth pathways and 9.9% (146 out of 1469) of its predictions were considered incorrect when comparing its predictions against the ground truth. To better understand these errors, we examined both the ground truth pathways missed and low-similarity pathway predictions.

First, analysis of missed ground truth pathways identified two primary causes. In most cases, the gene cluster detection process did not explicitly define gene sets that aligned with the functional terms in the ground truths. Additionally, in some instances, LLMs did not recognize the ground truth as the most representative biological function of the given gene cluster, despite its inclusion among the candidate enrichment results. These findings suggest that future improvements in gene cluster detection, for example, by refining functional modules within DEGs using gene embeddings, could enhance recall, thereby improving the accuracy and completeness of functional enrichment analysis.

Second, we manually reviewed 146 pathway predictions with low similarity scores (below 0.5) to ground truth pathways and found that 91.8% (134 out of 146) of predicted pathways were actually annotated as contextually relevant (**Data file S3, Low-similar Pathway Annotation**) by our expert annotators. For example, cGSA identified the "Inflammatory Response" as relevant for Alzheimer's disease and the "Adipogenesis" for rheumatoid arthritis, both of which were confirmed by domain experts as contextually appropriate. However, the highest similarity score between these pathways and their ground truths were only 0.469 and 0.422, respectively.

**Relevance Score Calibration**

The relevance scores assigned by cGSA to output pathways reflect the confidence of LLMs in assessing the contextual relevance of their results. Previous studies have shown that LLMs tend to exhibit overconfidence to its output, increasing the risk of generating incorrect contents, i.e., hallucinations[62,63,64]. To investigate this issue, we examined the correlation between relevance scores and two independent metrics: (1) the number of enrichment genes curated in databases (**Figure S3, a**), and (2) the semantic similarity scores calculated by MedCPT between cGSA's outputs and ground truths (**Figure S3b**).

The results indicate a linear-like upward trend, where both the number of enriched genes and similarity scores increase as relevance scores rise. On the one hand, the number of genes overlapping with curated enrichment results has long been a key metric for evaluating the significance of gene set enrichment analysis, such as in gene ontology annotation tasks. The observed linear trend in **Figure S3 (a)** indicates that higher relevance scores reflect greater functional relevance between predicted pathways and gene clusters. On the other hand, ground truths have been confirmed as contextually relevant in their original studies. Therefore, a higher similarity between cGSA's outputs and ground truths should generally correspond to higher contextual relevance. The positive correlation between relevance scores and similarity scores in **Figure S3 (b)** further supports this relationship, demonstrating that higher relevance scores correspond to a greater likelihood of pathways being contextually relevant.

These findings suggest that the relevance scores assigned by cGSA align well with independent similarity assessments, supporting their reliability in evaluating pathway-context relationships. Moreover, this alignment further demonstrates that the structured chain-of-thought prompt (**Data file S2**) designed in cGSA contributes to mitigating the overconfidence of LLMs, since it integrates three distinct criteria to assign a relevance score for each pathway: contextual relevance, pathway function analysis, and gene co-expression analysis.

We also examined the differences in relevance scores assigned to curated gene sets and their corresponding contaminated gene sets. Specifically, we collected 100 Gene Ontology (GO) terms with their associated gene sets ("curated gene sets"; **Figure S4a**) from *Ref.21.* Each GO term was then replaced with a synthetic gene set in which 50% of genes were randomly selected from the original GO term and 50% were randomly drawn from a background pool of all genes with GO annotations ("50% contaminated gene sets"; **Figure S4b**). Additionally, we evaluated a fully randomized variant in which 100% of genes were randomly selected from the background pool ("100% contaminated gene sets"; **Figure 4c**). The contextual descriptions for all evaluated gene sets were derived from the standard GO term definitions, as annotated by the QuickGO API (https://docs.identifiers.org/pages/api.html).

The results demonstrated that curated gene sets had an average relevance score of 8.31, which was higher than the 6.79 average score of the 50% contaminated gene sets and significantly greater than the 0.64 average score of the 100% contaminated gene sets. These findings further confirm that cGSA effectively captures the contextual relevance of gene sets by distinguishing between curated and contaminated sets based on their relevance scores. The substantial decrease in relevance scores observed in the 50% and 100% contaminated gene sets indicates that cGSA is sensitive to perturbations in gene composition. Specifically, the moderate decline in scores for the 50% contaminated gene sets suggests that introducing unrelated genes dilutes the contextual specificity of the original GO term, whereas the drastic reduction in scores for the 100% contaminated gene sets confirms that fully randomized gene sets lack meaningful biological coherence (**Data file S5**).

**The cGSA Online Application**

To the best of our knowledge, as the first and only gene set analysis method capable of incorporating contextual information, cGSA enhances researchers' ability to interpret and utilize high-throughput data in practical applications. By considering the biological context in which DEGs are produced, cGSA facilitates the identification of functional pathways and biological processes relevant to specific research objectives. This capability not only helps reduce the manual effort required to filter irrelevant results, but – more importantly – it also enables the discovery of disease-related pathways and previously unrecognized gene associations, as shown in the two independent case studies.

To streamline the user experience, cGSA is implemented as an end-to-end framework with optimized LLM prompts. Rigorous evaluations ensure the reliability of the framework, allowing users to simply input DEGs, the associated contexts, and optional parameters to obtain contextually relevant pathway predictions according to their study objectives. A demonstration website is freely available at https://www.ncbi.nlm.nih.gov/CBBresearch/Lu/Demo/cGSA/.

In addition to standard input requirements – such as a group of genes, enrichment database (default: KEGG pathway analysis), and species selection (default: human) – cGSA uniquely requires users to provide two types of context descriptions: (1) experimental conditions, specifying the experimental source of the gene set, and (2) research objectives, indicating the intended purpose of the gene set analysis. The results, including pathway summaries generated by cGSA, gene clusters, and enriched genes from the selected databases, are delivered via the email (see an example in **Data file S6**).

**Limitations and Future Directions**

While cGSA demonstrates superior performance in analyzing contextual relevance pathways, it may still generate pathways with low contextual relevance, potentially due to LLM hallucinations. Although structured prompt engineering helps mitigate this risk, further refinement of how predictions are influenced by different aspects of the context could improve accuracy. Our benchmarking focused on human and mouse DEGs, as these species are the most commonly used in GSEA databases. While they are the most studied, additional species could be explored in future work. Additionally, while our benchmark dataset currently comprises 102 DEG sets, establishing the first standardized benchmark for evaluating context-aware gene set analysis methods, it can be further expanded by integrating additional gene sets from publicly available databases in future work.

## MATERIALS AND METHODS

### Study Design

In this study, we developed cGSA to integrate clinical context into large language models for disease-specific pathway identification. cGSA is implemented using GPT-4o (version 20240513) via the Azure OpenAI API, a HIPAA-compliant platform[65] that ensures robust data privacy protections. All model queries were executed with a temperature of 0.0 to guarantee deterministic and reproducible outputs.

We evaluated cGSA using 102 DEG sets (mean length is 914 genes) curated from 31 disease-related articles, which were manually selected from 331 PubMed articles published between October 2022 and June 2024. Article inclusion criteria required the presence of clearly defined research objectives, experimental conditions, pathway annotations, and DEG lists containing at least 50 genes. All gene sets were normalized as the standard NCBI gene symbols and preprocessed through the STRING API and EdMot algorithm to produce biologically coherent gene clusters, ensuring that evaluated inputs remained novel relative to GPT-4o's pretraining corpus. Each gene set was analyzed once using fixed parameters, with no randomization or stochastic sampling.

We further tested cGSA on nine independent DEG sets: six derived from preclinical melanoma models (mean length is 758 genes) and three from breast cancer studies (mean length is 107 genes). These data were provided by collaborators based on previously published or ongoing research projects. Only finalized DEG lists were received, without access to the original experimental datasets.

### Statistical analysis

All statistical analyses were performed using Python (version 3.11.0). The cGSA framework was implemented using requests (v2.32.2) and openai (v0.28.0). The STRING and Enrichr APIs used were versions 12.0 and 20230608, respectively. For evaluation, we developed an automated benchmarking pipeline using torch (v1.13.0) and transformers (v4.43.3), incorporating MedCPT (version 20231025) to compute similarity scores between cGSA-generated outputs and expert-defined ground truths. Additional packages included numpy (v1.26.3) and pandas (v2.1.4). Performance metrics included accuracy, hit ratio, and harmonic F1 score, with results visualized using seaborn (v0.13.2). All figure results are presented as mean plus the standard deviation (SD), with sample sizes indicated in figure captions.

For baseline comparisons, we included GPT-4 (version 20230613) and GPT-3.5 (version 20230613), both accessed via the Azure OpenAI API with temperature set to 0.0. We also evaluated Llama 3.1 70B Instruct (version 20240418, knowledge cutoff: December 2023), accessed through the Hugging Face transformers interface[66] with sampling disabled to ensure deterministic outputs. Additionally, traditional gene set enrichment was conducted using g:Profiler (API version e112_eg59_p19_25aa4782), with a significance threshold of $p < 0.05$. Statistical significance for cGSA's performance improvement was assessed using one-tailed paired *t*-tests at a 95% confidence level.

In addition to automatic evaluation, two independent domain experts performed blinded annotation of the outputs. Inter-annotator agreement was calculated as absolute consistency between the annotations.

**cGSA workflow**

The cGSA workflow consists of four integrated steps, each designed to address specific challenges in GSA and is illustrated in **Figure 1b**: (1) *Gene Cluster Detection*: Using community detection algorithms applied to gene networks derived from the STRING database, cGSA identifies functional gene clusters based on protein-protein interactions (PPIs). (2) *Gene Cluster Enrichment*

*Analysis*: Each gene cluster is analyzed with conventional gene set enrichment analysis using user-specified databases (e.g., KEGG, Reactome) through the Enrichr API. (3) *Pathway Screening with LLMs*: To incorporate experimental context, large language models (LLMs) are leveraged to refine and prioritize enriched pathways. (4) *Pathway Summarization with LLMs*: To improve interpretability, cGSA employs summarization-based instructions for LLMs, condensing the refined list of pathways into concise, contextual coherent outputs.

Together, these steps enable cGSA to deliver biologically meaningful and contextually relevant pathway analyses, addressing the fragmented and irrelevant outputs of traditional enrichment analysis. The entire workflow is summarized in **Figure 1**, and each step is described in detail as follows.

(1) *Gene cluster detection*. The first step identifies functional groups of genes within a protein-protein interaction (PPI) network by clustering the input gene set, which can be derived from various data sources such as DEGs or other high-throughput experiments, into non-overlapping communities. These clusters capture biologically relevant patterns, such as co-regulated or functionally related gene modules[67], which are often missed by traditional GSEA methods.

Given a set of genes ($\mathcal{D}$) as an input, we first utilized the STRING API (https://string-db.org/help/api/)[33] to construct the gene interaction network represented as $\mathcal{G} = (\mathcal{V}, \mathcal{E})$, where $\mathcal{V}$ denotes a subset of genes in $\mathcal{D}$, and $\mathcal{E}$ represents interactions between two genes ($v_i, v_j \in \mathcal{V}$) with the confidence score greater than 0.05. To handle large gene sets exceeding the STRING API limitation of 2,000 genes, we implemented a batch strategy. In this strategy, we queried 1,000 genes at a time, with 200 genes overlapping between consecutive queries to ensure continuity and minimize missing interactions.

Once the gene interaction network $\mathcal{G} = (\mathcal{V}, \mathcal{E})$ has been constructed, we applied the EdMot algorithm[34] to analyze the network. The algorithm divides $\mathcal{V}$ into non-overlapping clusters (or communities) $(c_i)$ capturing distinct functional patterns within the network. This cluster process can be formally represented as:

$$\mathcal{C} = \bigcup_{i=1}^{K} c_i\,(v_i, \varepsilon_i) = EdMot(\mathcal{G}, \Theta) \tag{1}$$

Here, $K$ is the total number of clusters, and $\Theta$ represents the hyperparameters of the EdMot, which controls the number of components and cutoff thresholds for each cluster. More details can be referred to *Ref.34*.

(2) *Gene cluster enrichment analysis*. Building on the clustering step, this process identifies statistically enriched pathways for each gene cluster, forming the basis for contextual refinement in subsequent steps. Each gene cluster $c_i(v_i, \varepsilon_i)$ obtained from the previous step, is analyzed using the user-specified databases via the Enrichr API (https://maayanlab.cloud/Enrichr)[18]. Enrichment analysis is performed on $v_i$ (the genes within each cluster) using curated databases $(db)$ to identify significantly enriched pathways. After the enrichment analysis, each $c_i$ is assigned a set of enrichment terms $(\mathfrak{T})$, such as pathway, associated with a significant p-value $(p)$. This process can be formally represented as:

$$c_i^{\mathfrak{T}} = Enrichr(c_i, p; db) \tag{2}$$

The threshold cutoff of p-value $(p)$ is 0.001 according to experimental results. The selected $db$ was adjusted to align with the experimental conditions and chosen from the backend databases provided by Enrichr.

(3) *Pathway screening with LLMs*. This step integrates contextual information, such as treatment conditions and enrichment of interest, to ensure that pathway analysis aligns with the specific experimental settings. Since the enrichment pathways ($c_i^{\mathfrak{T}}$) for each gene cluster can be extensive, we incorporated the experimental conditions ($C_E$) used to identify the regulated gene in the original research to reduce the scale of $c_i^{\mathfrak{T}}$.

To achieve this, we designed prompts for LLMs to access the relevance of each pathway in $c_i^{\mathfrak{T}}$ with respect to $C_E$, based on the functional analysis of pathways and their enriched genes in the database. This relevance scoring process can be formally expressed as:

$$\mathcal{S}_{c_i^{\mathfrak{T}}} = LLM(c_i^{\mathfrak{T}}, prompt^1; C_E) \quad (3)$$

$\mathcal{S}_{c_i^{\mathfrak{T}}}$ represents the relevance scores assigned to pathways within $c_i^{\mathfrak{T}}$ and $prompt^1$ contains the contextual instructions provided to the LLM for this analysis. Detailed information about $prompt^1$ is provided in **Data file S2**.

(4) *Pathway summarization using LLMs*. To minimize the hallucination risks associated with LLMs, the step prioritizes the most representative pathways based on the given research objectives. After obtaining the filtered candidate pathways $\mathcal{S}_{c_i^{\mathfrak{T}}}$, all significant pathways within a gene cluster are aggregated and input into an LLM prompt for summarization. The goal is to derive the most representative pathway ($\mathcal{P}_{c_i^{\mathfrak{T}}}$) based on the research objectives ($C_R$). Here, the $C_R$ provides the contextual background knowledge specific to the study, ensuring that cGSA aligns with the researchers' requirements through the analysis of the input gene sets. Additionally, inspired by the prior research on gene set analysis by Hu *et al.*[21], we developed an instruction within the LLM prompt to assign a confidence score ($\mathcal{R}$) to the summarized pathway. This score reflects the pathway's relevance to the given context, offering an additional layer of interpretability. This summarization process can be formally expressed as:

$$\mathcal{P}_{c_i^{\mathfrak{T}}}; \mathcal{R} = LLM(\mathcal{S}_{c_i^{\mathfrak{T}}}, prompt^2; C_R) \tag{4}$$

The $prompt^2$ contains specific instructions guiding the LLM to generate the summary and confidence score. Details about $prompt^2$ are provided in **Data file S2.**

In summary, the pseudocode of cGSA can be presented as follows:

```
# Constructing the gene interaction network 𝒢 = (𝒱, ℰ) for DEGs (𝒟)
if gene numbers in 𝒟 is less than 2000:
    Construct 𝒢 using the STRING database.
else:
    Construct 𝒢 using batch sampling and the STRING database.
# Gene cluster detection
Apply Eq. (1) on 𝒢 to partition it into 𝒞 = ⋃_{i=1}^{K} c_i (v_i, ε_i)
for c_i in 𝒞:
    # Gene cluster enrichment analysis
    Apply Eq. (2) on c_i to obtain all significant pathways c_i^𝔗
    # Pathway screening
    Apply Eq. (3) on c_i^𝔗 to screen the context-related pathways 𝒮_{c_i^𝔗}
    # Pathway summarization
    Apply Eq. (4) on 𝒮_{c_i^𝔗} to present the representative pathway 𝒫_{c_i^𝔗} (with ℛ)
end for
```

**Data Collection**

Scientific literature provides a wealth of research containing biomedical hypotheses supported by gene set enrichment analysis. Such pathway and functional signatures of gene sets are manually reviewed and recorded in studies, with gene sets frequently included in supplementary materials[68]. For example, a study (PMID39241064)[69] focuses on the role of threonine-4 (Thr4) phosphorylation on the C-terminal domain (CTD) of RNA polymerase II, highlighting its significance in transcription regulation and 3'-end processing of mRNA. The paper specifically addresses the functional consequences of Thr4 phosphorylation in RNA Polymerase II, emphasizing its role in transcription termination and 3'-end processing. It further illustrates the essential biological processes of negatively regulated genes upon 52X T4A-CTD expression, which were manually selected from hundreds of GO terms identified through GSEA. Accompanying the key biological processes, the article supplements the DEG sets in supplementary table S3 of the paper. As it describes both the hypothesis and its demonstration of biological meaning, it serves as a robust dataset to evaluate our method.

Drawing from this example, the data collected for evaluation of cGSA includes four key factors: (1) Research objectives: "Impact of Threonine 4 phosphorylation of RNA Pol II on gene expression"; (2) Experimental conditions: "negatively regulated genes upon 52X T4A-CTD expression"; (3) Selected functions/pathways, which are usually presented in figures or tables in the main text; and (4) DEG sets, which is usually archived in the supplementary tables.

For the general evaluated data, we queried PubMed using the search term "Gene Set Enrichment Analysis (GSEA)" and restricted the publication timeline to between October 2022 and June 2024. The scope was further narrowed to articles in the PMC Open Access Subset or those including supplementary tables. We then applied GNorm2[70] to recognize gene mentions in the supplementary tables and count their frequencies. Only articles with at least 50 genes in the supplementary tables were selected. In total, 331 articles met the criteria and were available for manual curation. Over four weeks, we carefully reviewed each DEG set along with the corresponding research objectives, experimental conditions, and manually selected functions or pathways described in each article. Ultimately, we curated 102 DEG sets from 31 articles as ground truths, and this evaluation dataset was archived in **Data file S1, Evaluated DEGs**.

For the case studies, all data were provided by our collaborators and originated from their previously published work or ongoing research projects. We received only the final processed DEG lists for analysis and did not have access to the original raw experimental data. As such, our analyses were limited to the curated gene lists and did not involve reprocessing of primary experimental measurements. This dataset was archived in **Data file S1, Case-Study DEGs**.

**Automatic Evaluations Using MedCPT.**

Each gene set ($\mathcal{D}$) was annotated with several ground truths ($\mathcal{Y}$) confirmed by the original authors. To access the similarity between the pathways summarized by cGSA and the corresponding ground truths, we first encoded all pathways $\bigcup_{i=1}^{K} \mathcal{P}_{c_i^{\mathfrak{X}}}$ proposed by cGSA and $\mathcal{Y}$ as the embeddings using the MedCPT text encoder[41]. Next, we calculated the cosine similarity score between each paired vector ($\mathcal{P}_{c_i^{\mathfrak{X}}}, y \in \mathcal{Y}$). We applied various similarity score thresholds (ranging from 0.5 to 0.9) to classify cases as positive or negative. A many-to-many calculation strategy were employed, where a pathway $\mathcal{P}_{c_i^{\mathfrak{X}}}$ was considered meaningful if its similarity score with at least one ground truth exceeding the threshold.

The evaluation metrics are selected as accuracy (ACC), hit ratio (HIT), and their harmonic mean. The HIT represents the proportion of ground truths covered by the cGSA's output pathways, while ACC denotes the proportion of pathways aligned with the ground truths in all output pathways. These metrics can be formally expressed as:

$$HIT = \frac{\left| numbers\ of\ ground\ truths\ covered\ by\ \bigcup_{i=1}^{K} \mathcal{P}_{c_i^{\mathfrak{X}}} \right|}{|\mathcal{Y}|} \tag{5}$$

$$ACC = \frac{\left| numbers\ of\ meaningful\ pathways\ in\ \bigcup_{i=1}^{K} \mathcal{P}_{c_i^{\mathfrak{X}}} \right|}{K} \tag{6}$$

$$Harmonic\ Mean = (2 * HIT * ACC)/(HIT + ACC) \quad (7)$$

**Human Validations by Biologists**

To validate the effectiveness of the automatic evaluation based on similarity scores, two biologists annotated functional categories for the output pathways ($\bigcup_{i=1}^{K} \mathcal{P}_{c_i^{\mathfrak{T}}}$) of each gene cluster ($c_i$) as well as for each ground truth (**Data file S3, Functional Category Annotation**). Output pathways that matched the functional category of the ground truths were designated as function-related pathways. Using these function-related pathways, we then calculated accuracy (ACC) and hit ratio (HIT) to assess performance. HIT was defined as the proportion of ground truths covered by function-related pathways, while ACC represented the proportion of function-related pathways among all output pathways. To provide a balanced measure of performance, we computed the harmonic mean of ACC and HIT (**Figure 3**).

**Acknowledgments:** We would like to thank M.G. Hirsch and Teresa M. Przytycka for their helpful discussion of this work.

**Funding:** This research is supported by the NIH Intramural Research Program, National Library of Medicine (NLM) and National Cancer Institute (NCI).

**Author contributions:** Z.W., C.D., Q.J., and Z.L. conceived this study. Z.W. and C.W. implemented the data collection and quality control. C.D. provided the gene sets derived from the mouse B2905 melanoma cell line. A.Q. and X.L. provided the gene sets derived from the breast cancer. Z.W. and Y.Y. conducted the model construction and evaluation. C.D. and A.Q. carried out the human validation of the main results. Z.W., Y.F., and Q.Z. performed the human inspections for PubMed articles. Z.W., C.W., R.L., C.D., Y.Y., A.Q., Y.F., X.L., and Z.L. conducted manuscript drafting & revision. C.W., S.T., and Z.W. developed the website of cGSA. Z.L. supervised the study. All authors approved the submitted version.

**Competing interests:** All authors declare no competing interests.

**Data and materials availability:** All evaluated differentially expressed genes and gene sets were collected from publicly available articles identified through PubMed searches. These gene sets, along with those used in the case study, are provided in **Data file S1** accompanying this submission. The biomedical text encoder MedCPT is available at https://github.com/ncbi/MedCPT. We provide a brief preview of the cGSA code at: https://gist.github.com/AllenWangle/56ef9e75fa6d8f69d19f864b232d3435. A website for cGSA is available at: https://www.ncbi.nlm.nih.gov/CBBresearch/Lu/Demo/cGSA/

## Figures and Tables

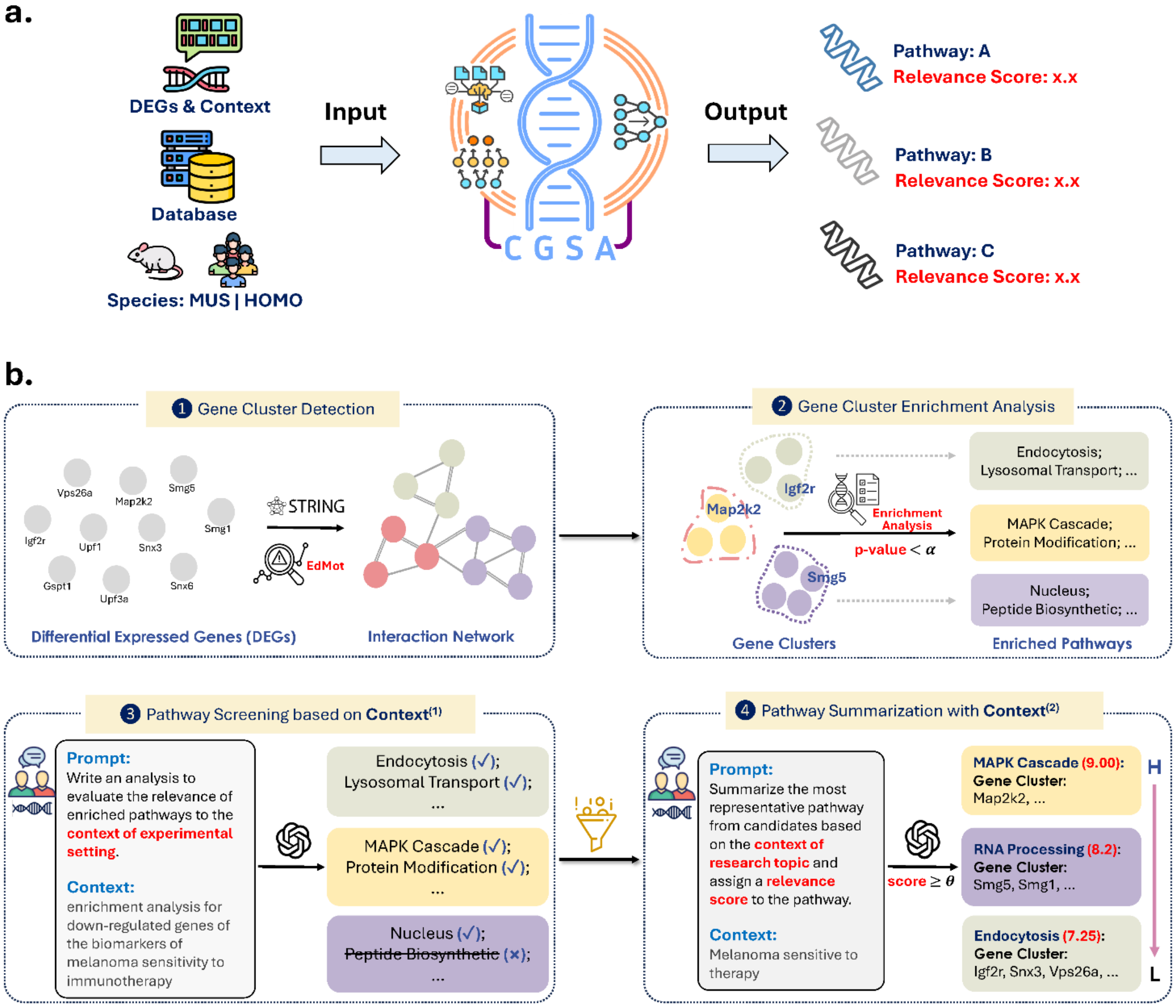

**Fig. 1. The cGSA workflow, which integrates multiple computational methods to identify contextually relevant pathways from input gene sets. a**, overview of the cGSA system. **b**, the cGSA pipeline consists of four key steps. *Gene Cluster Detection* groups input gene set into distinct communities using the STRING database and the EdMot algorithm. *Gene Cluster Enrichment Analysis* identifies the most significant pathways for each gene cluster (p-value<0.001) by querying established target databases. *Pathway Screening* employs large language models (LLMs) to filter enriched pathways based on the experimental or biological context. *Pathway Summarization* selects the most representative pathway for each gene cluster by integrating the filtered pathways with the broader research context.

The final output pathways are ranked by their relevance scores, from high (**H**) to low (**L**), based on relevance scores (score > 5.5; **Figure 3c, 3d**). The detailed LLM prompts are provided in **Data file S2.**

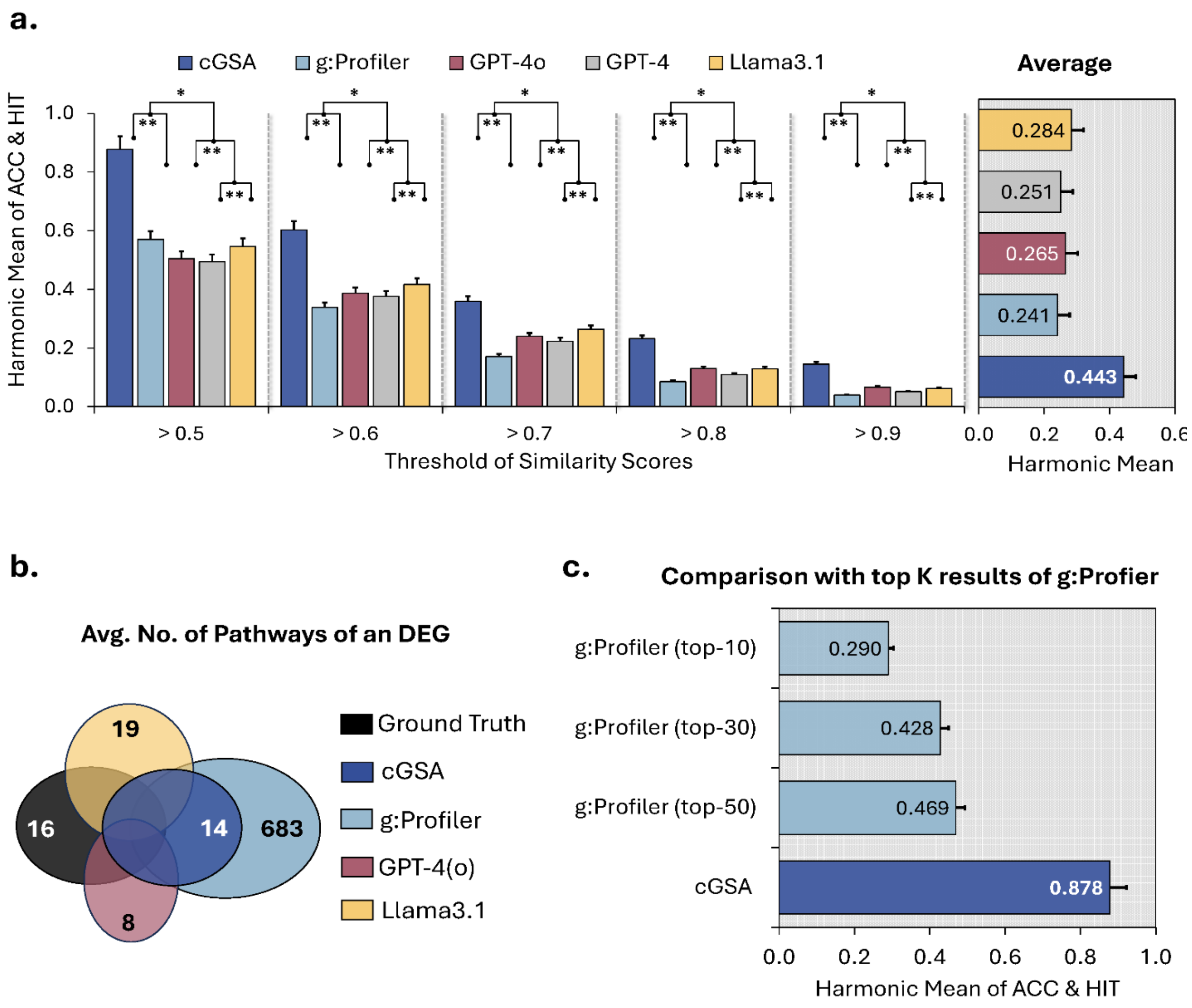

**Fig. 2. cGSA outperforms existing approaches by a significant margin.** **a**, the harmonic means of accuracy (ACC) and hit ratio (HIT) achieved by cGSA and other baseline methods across different similarity score thresholds. *HIT* represents the proportion of ground truths covered by the cGSA's output pathways, while *ACC* denotes the proportion of pathways aligned with the ground truths in all output pathways. "**" indicates a significant improvement with $p < 0.01$, while "*" denotes a significant improvement with $p < 0.05$. The p-values are calculated by a one-tailed paired T-test with 95% confidence intervals for 102 DEGs. The average values across all thresholds are presented in the right panel with the error bar indicating the standard deviation. **b**, the average number of pathways

identified by different methods for each input DEGs. GPT-4(o) indicates that both GPT-4 and GPT-4o produced the same average number of pathways per DEGs. **c**, performance comparison between cGSA and the top K (K = 10, 30, 50) results from conventional gene list enrichment analysis (g:Profiler) with the similarity score threshold of 0.5. The error bar denotes the standard errors.

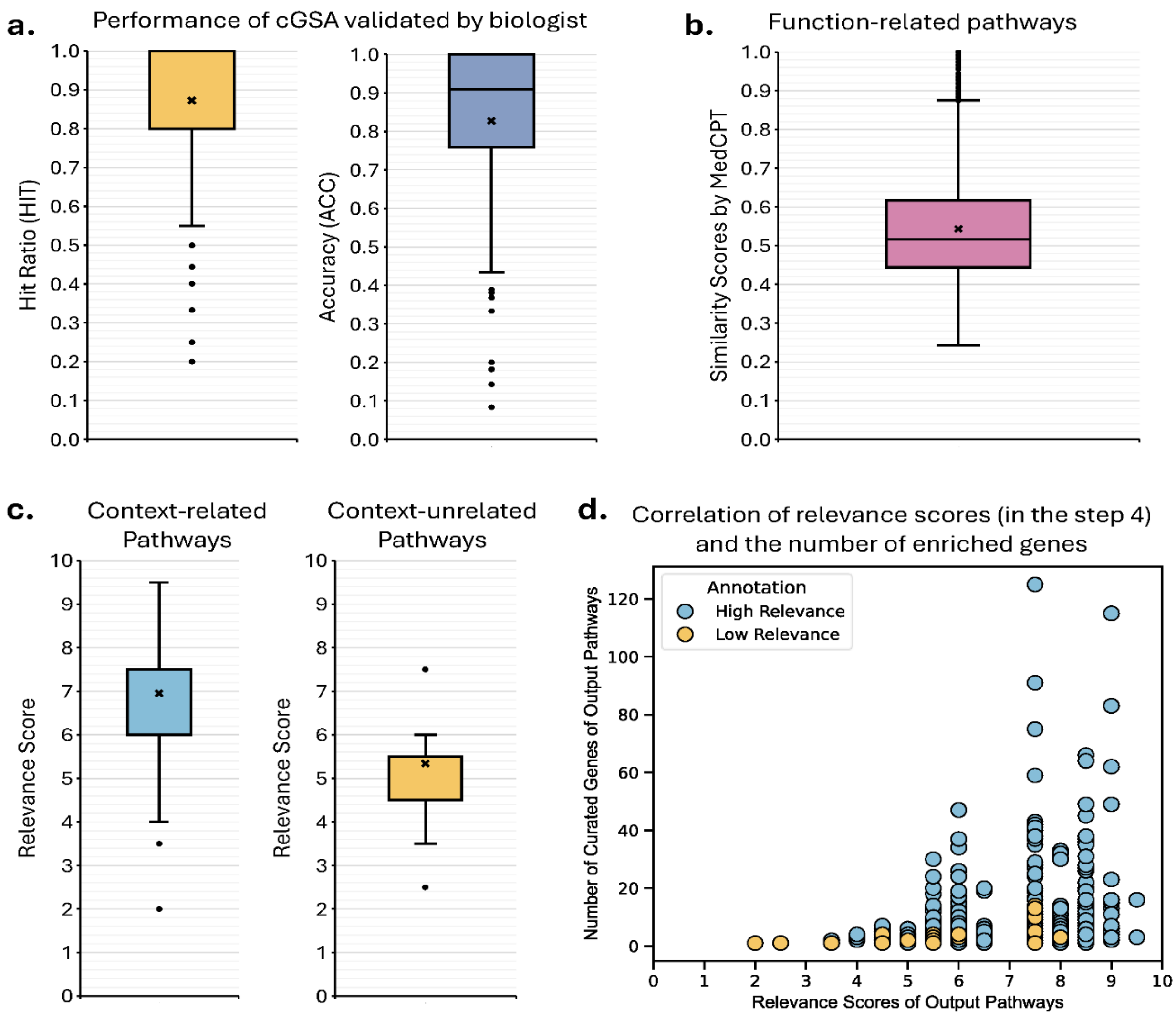

**Fig. 3. Annotation analysis of pathways identified by cGSA. a,** the calculation of hit ratio (HIT) and accuracy (ACC) for output pathways (n=1,469) with the same functional category as the ground truths (n=1,671), i.e., function-related pathways. *HIT* represents the proportion of ground truths covered by function-related pathways, while *ACC* denotes the proportion of function-related pathways in all identified pathways. **b**, the distribution of semantic similarity scores for function-related pathways (n=1,216), as computed by MedCPT. **c**, the relevance scores of function-related pathways (n=1,216) compared to function-unrelated pathways (n=253) to the context, where the context refers to the research topic. **d**, correlation between relevance scores of identified pathways (n=1,469) and the number of their enriched genes curated in the database. For boxplots, the lower and upper hinges correspond to the first and third quartiles. The upper (lower) whisker extends from the hinge to the largest (smallest) value no further than (at most) 1.5× IQR from the hinge (where IQR is the inter-quartile range). The middle points represent the mean values while the middle lines represent the median values.

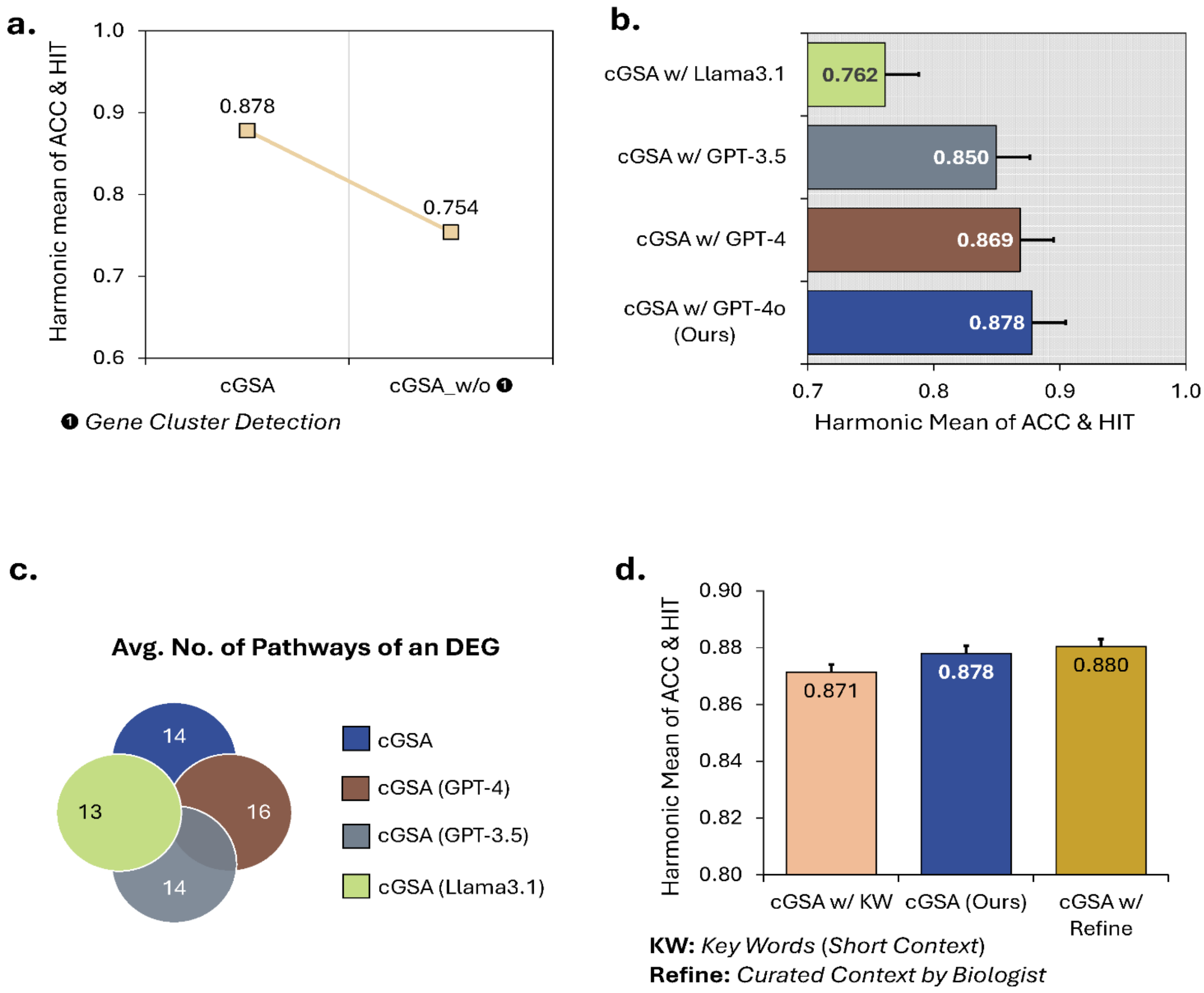

**Fig. 4. Ablation experiments for cGSA with a similarity score threshold of 0.5. a,** comparison between cGSA and its variant without the Gene Cluster Detection step (cGSA w/o ❶) **b,** performance comparison of cGSA using different large language models (LLMs). **c,** the average number of output pathways per DEG identified by cGSA using different LLMs. **d,** performance comparison of cGSA across different research topic contexts. "KW" refers to keywords (2.59 averaged words) summarized from the context used in cGSA, while "Refine" represents the context manually curated by the biologist (8.71 averaged words) based on the original context used in cGSA. The error bars indicate the standard errors.

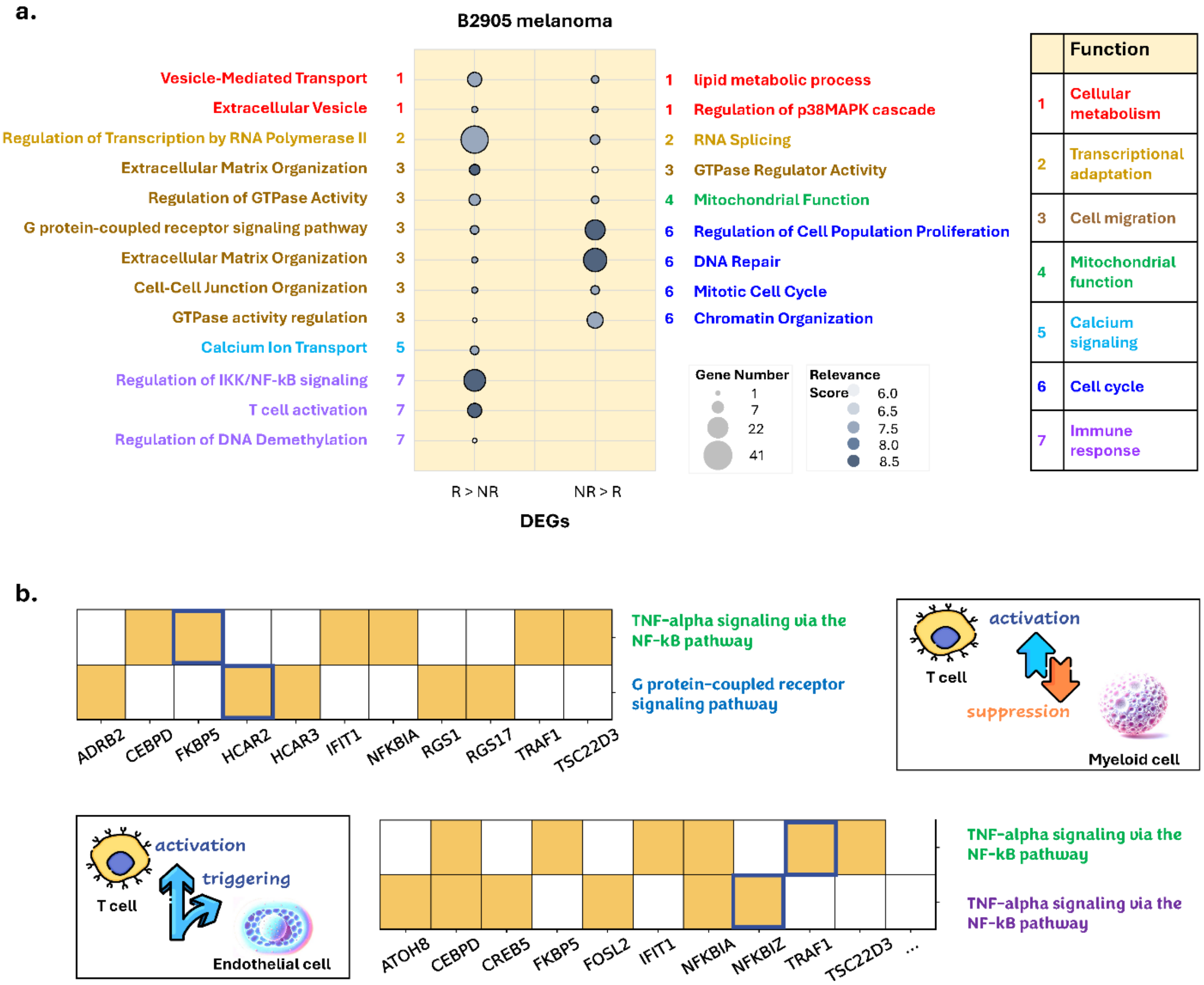

**Fig. 5. Case studies of melanoma cancer and breast cancer. a,** cGSA-identified pathways enriched by the differentially expressed genes (DEGs) between the responding (R) and non-responding (NR) B2905 melanoma treated with anti-CTLA-4 immune checkpoint blockade (ICB) therapy. R > NR refers to genes more highly expressed in R tumors compared to NR tumors. NR > R refers to genes more highly expressed in NR tumors to in R tumors. The numbers and color codes of functional pathway categories are provided in the accompanying table on the right. **b,** examples of pathways and genes summarized by cGSA for DEGs identified in T cells, myeloid cells, and endothelial cells to investigate cell-cell communication (CCC) within the tumor microenvironment (TME) of breast cancers. The highlighted genes are identified as biomarkers in specific pathways based on the researchers' causal knowledge. For instance, the activation of *FKBP5* in the T cell suppresses *HCAR2* in the myeloid cell (**top**). The activation of *TRAF1* in the T cell induces the activation of *NFKBIZ* in the endothelial cell (**bottom**).

**Table 1. Differentially expressed genes (DEGs) data set used to evaluate cGSA**.

| DEGs | #DEGs | Avg. #Genes | #Disease | #Mechanisms | Species | Resource |
|---|---|---|---|---|---|---|
| PMDEGs | 102 | 913.6 | 19 | 10 | Human & Mouse | PubMed articles |
| Melanoma Cell Lines | 6 | 757.5 | 1 | - | Mouse | *Ref.* 32 |
| Breast Cancer | 3 | 107.3 | - | 1 | Human | Published Data |

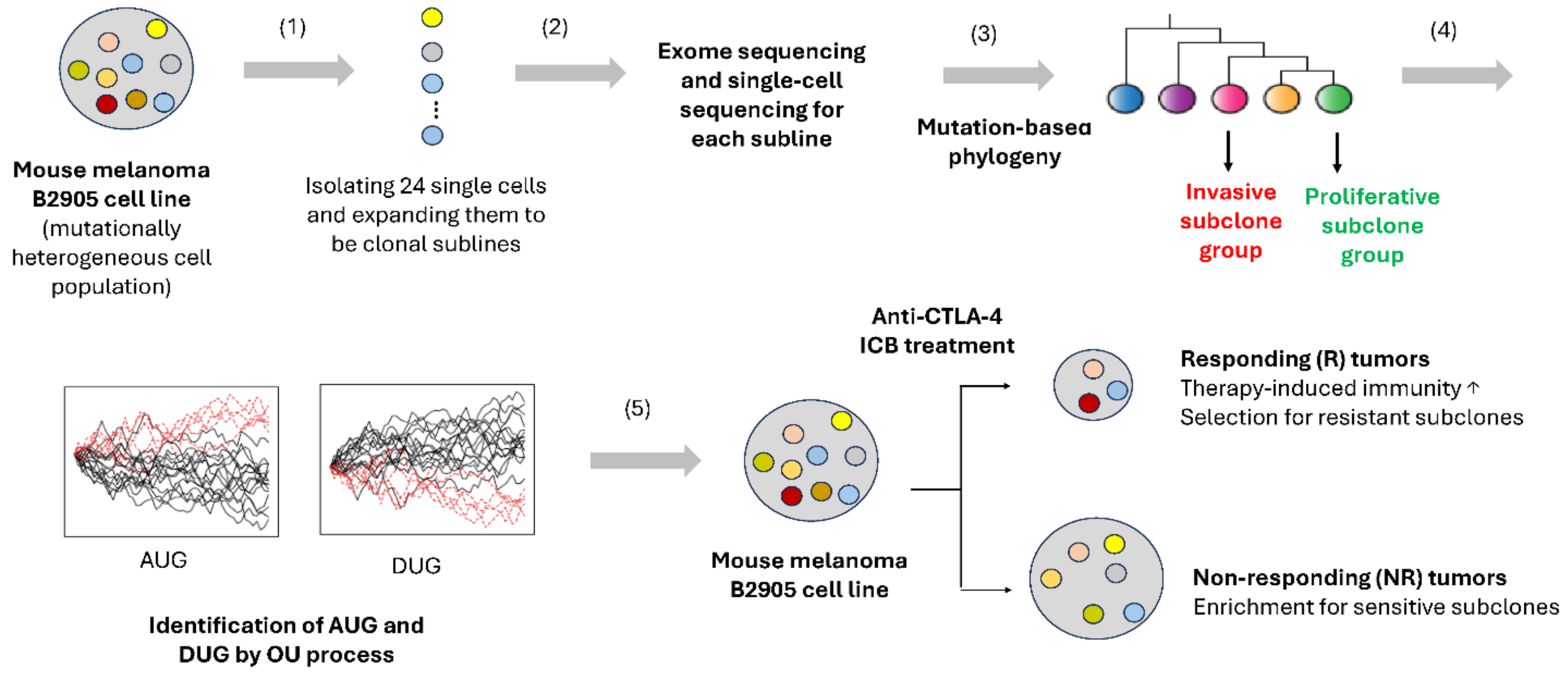

**Figure S1. The pipeline proposed by Hirsch *et al.* to model evolutionary changes in gene expression of melanoma as stochastic Ornstein-Uhlenbeck (OU) processes.** Briefly, 24 clonal sublines derived from the parental mouse melanoma B2905 cell line were analyzed for DNA mutations and RNA expressions. They built mutation-based phylogeny of the 24 sublines and characterized branches containing several evolutionarily related sublines, i.e. "subclone groups", whose sublines exhibited similar phenotypes. Two subclone groups were named "invasive subclone group" and "proliferative subclone group", based on their distinct phenotypes. Moreover, the former was resistant, and the latter was sensitive to immune checkpoint blockade (ICB) therapies. By analyzing evolutionary trajectories of gene expression along the phylogeny with OU process, they identified adaptively upregulated and downregulated genes (AUGs and DUGs, respectively) in the invasive (572 AUGs and 347 DUGs) and proliferative subclonal groups (169 AUGs and 1151 DUGs) (**Data file S1, Case-Study DEGs**).

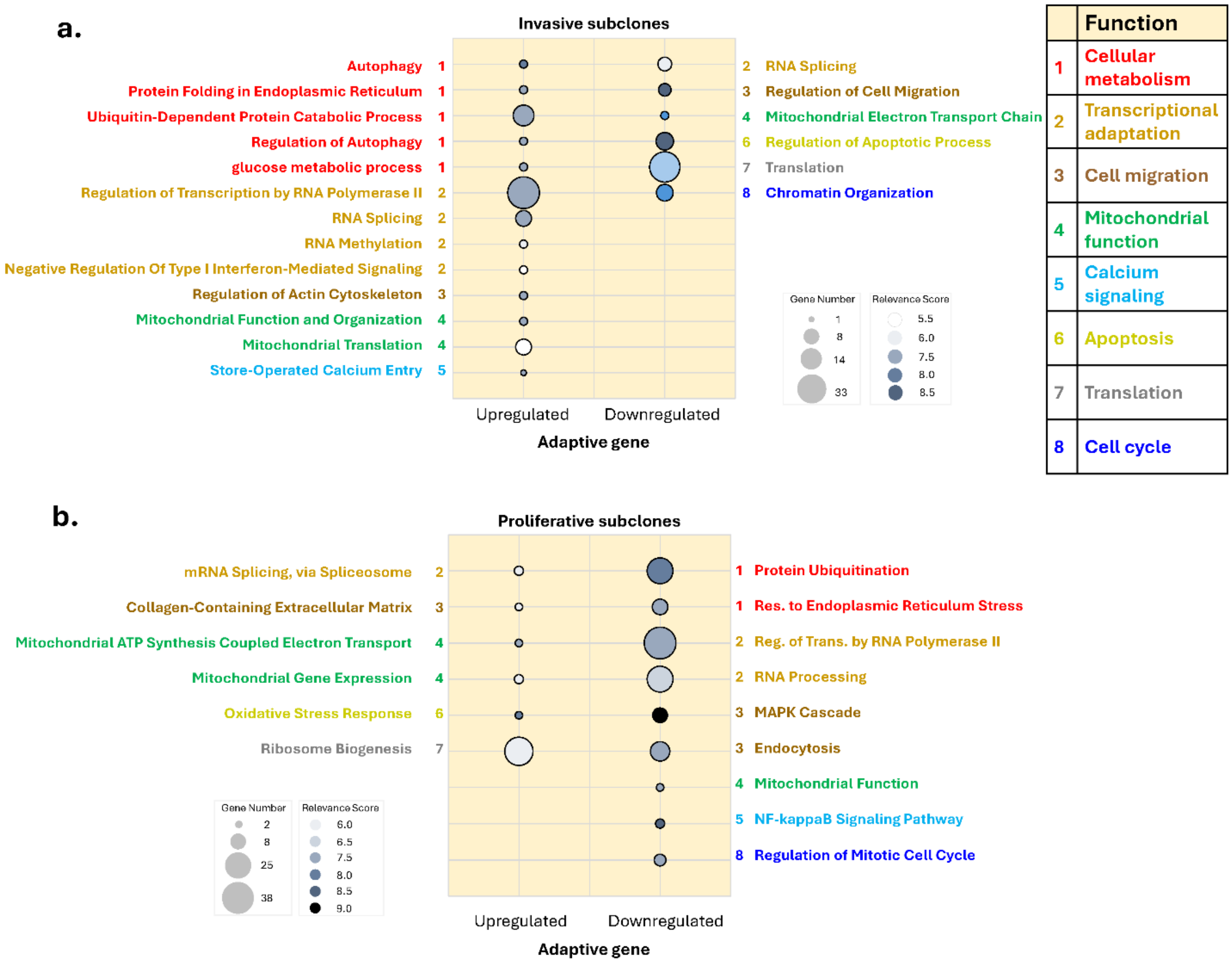

**Figure S2. cGSA output pathways enriched by the adaptively upregulated genes (AUGs) and adaptively downregulated genes (DUGs) of the invasive subclone group (a) and the proliferative subclone group (b).** The context is "melanoma". The results show that: (1) for the invasive subclonal group, AUGs were more enriched in category 1 (metabolism) and 2 (transcriptional adaptation) pathways with higher relevance scores. These include pathways of "Ubiquitin-dependent protein catabolic process" (14 genes, score = 7.5), "Regulation of transcription by RNA Polymerase II" (33 genes, score= 7.5), and "RNA splicing" (8 genes, score = 7.5). In contrast, DUGs of the invasive subclonal group were more enriched in category 6 (apoptosis), 7 (translation), and 8 (cell cycle) pathways with higher relevance score. These include pathways of "Regulation of apoptotic process" (10 genes, score = 8.5), "Translation" (30 genes, score = 7.0), and "Chromatin organization" (9 genes, score = 7.5). (2) for the proliferative subclonal group, AUGs were more enriched in category 6 (apoptosis) and 7 (translation) pathways with higher relevance score. These include pathways of "Oxidative stress response" (2 genes, score = 8.0), and "Ribosome biogenesis" (30 genes, score = 6.5). On the other hand, DUGs were more enriched in category 1 (metabolism), 2

(transcriptional adaptation), and 3 (cell migration) pathways with higher relevance score. These include pathways of "Protein Ubiquitination" (25 genes, score = 8.5), "Regulation of Transcription by RNA Polymerase II" (38 genes, score = 7.5), "RNA Processing" (25 genes, score = 7.0), "MAPK pathway" (8 genes, score = 9.0), and "Endocytosis" (14 genes, score = 7.5). In summary, in consistency with the observation by Hirsch *et al*, our results also demonstrate that the adaptive genes of invasive and proliferative groups show contrast functional patterns. That is, the functions referred to the adaptively up- and down-regulated genes in the invasive subclones are those to the adaptively down- and up-regulated genes in the proliferative subclones, respectively. This suggests that the two subclonal groups gained invasive and proliferative phenotypes by evolving toward two distinct transcriptional states, conferring the functions of metabolic/transcriptional adaptation and cell growth, respectively.

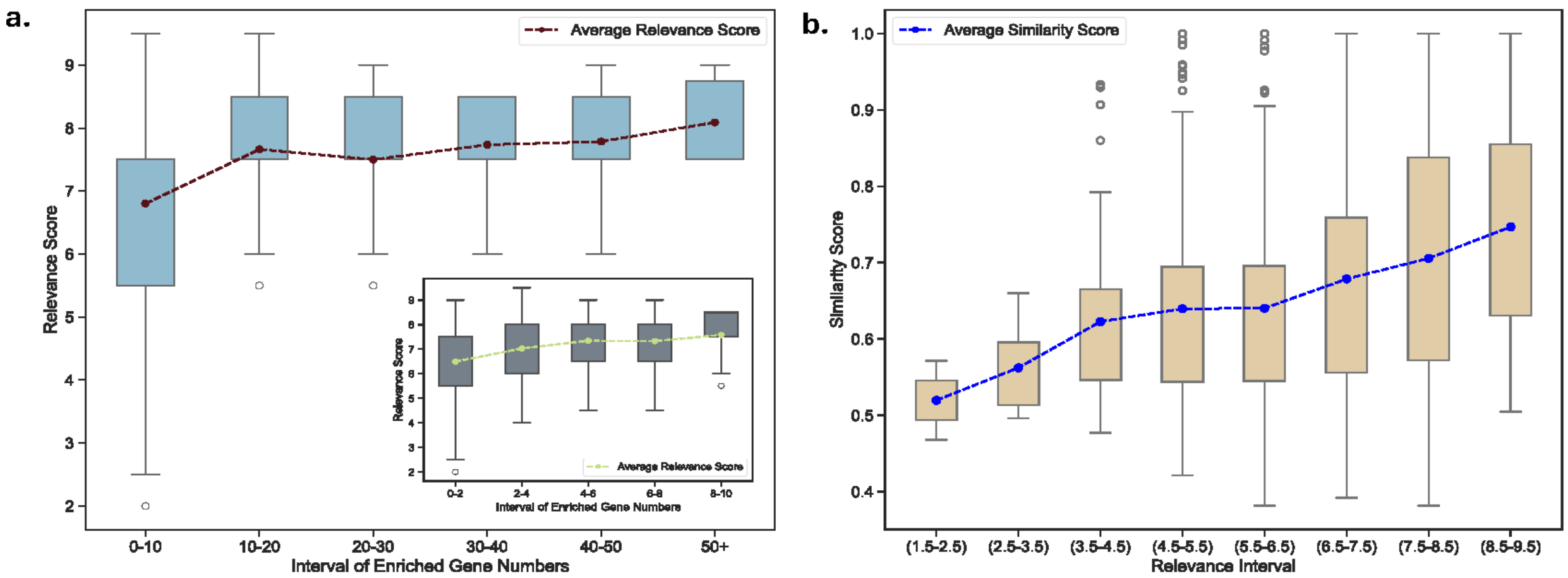

**Figure S3. The results of relevance score calibration.** Correlations between relevance scores generated by cGSA and the number of enriched genes of ground truths curated in databases (a), and the similarity scores calculated by MedCPT between cGSA's output and ground truths. We grouped the number of enriched genes (**a**) and relevance scores of all output pathways (**b**) into different intervals (**x-axis**). The average relevance score (**a**) and similarity score (**b**) in each interval (**y-axis**) between output pathways and their corresponding ground truths, exhibits a linear-like increasing trend. The total samples for statistical analysis are 1,469. For boxplots, the lower and upper hinges correspond to the first and third quartiles. The upper (lower) whisker extends from the hinge to the largest (smallest) value no further than (at most)1.5× IQR from the hinge (where IQR is the inter-quartile range). The middle points represent the mean values.

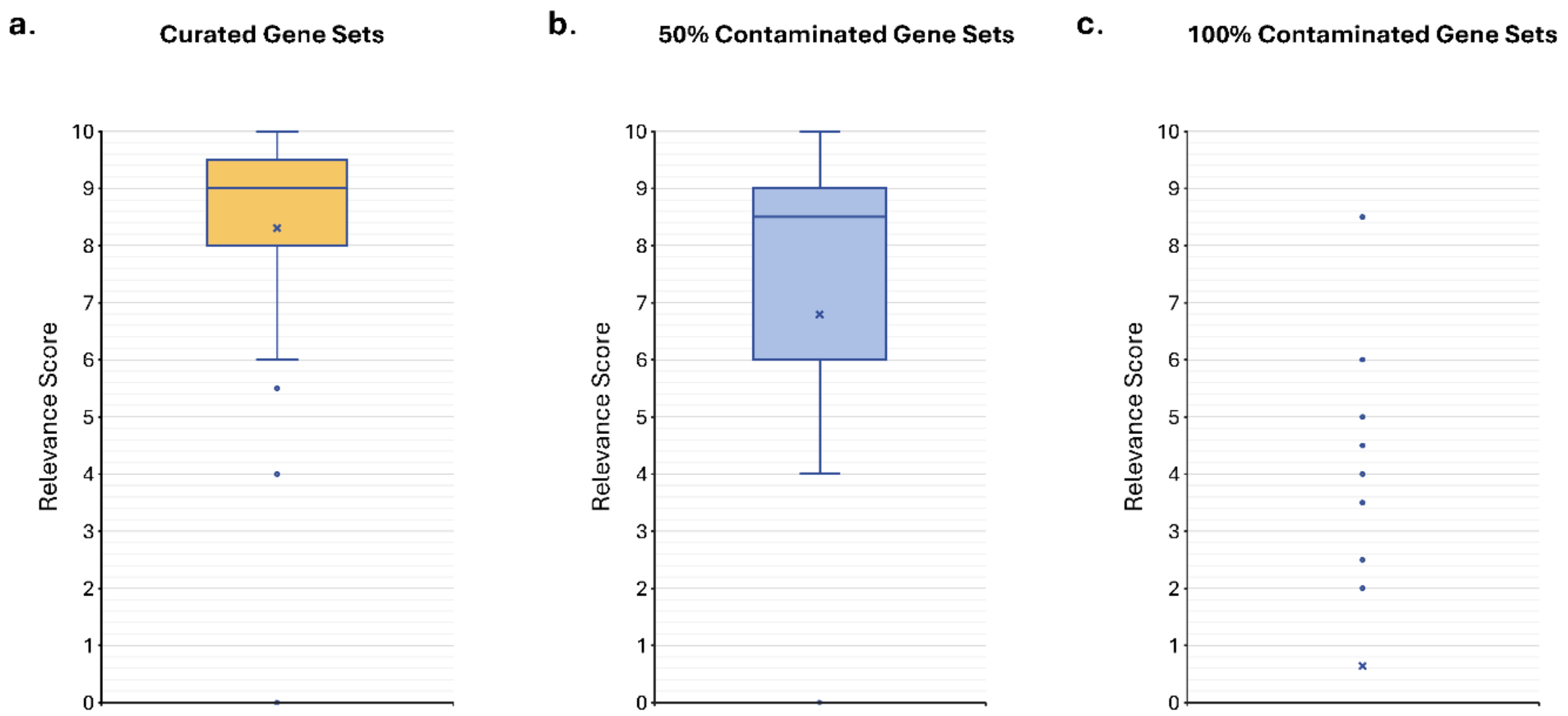

**Figure S4. Relevance scores for curated gene sets and their corresponding contaminated gene sets. a**, relevance scores assigned to curated gene sets containing no randomly introduced genes, with an average score of 8.31. **b**, relevance scores assigned to 50% contaminated gene sets, in which half of the genes were randomly selected from the background pool. The average score is 6.79. **c**, relevance scores assigned to 100% contaminated gene sets, where all genes were randomly selected while maintaining the same scale as curated gene sets. The average score is 0.64. The total samples for statistical analysis are 100. For boxplots, the lower and upper hinges correspond to the first and third quartiles. The upper (lower) whisker extends from the hinge to the largest (smallest) value no further than (at most)1.5× IQR from the hinge (where IQR is the inter-quartile range). The middle points represent the mean values while the middle lines represent the median values.

**List of Supplementary Materials**

Data file S1 (.xlsx): Differentially expressed genes (DEGs) used for evaluation and case studies.

Data file S2 (.docx): Prompts (instructions) used in the cGSA framework.

Data file S3 (.xlsx): Expert annotations for the outputs of cGSA.

Data file S4 (.xlsx): Results of two case studies.

Data file S5 (.xlsx): Evaluations of relevance scores for different DEGs.

Data file S6 (.xlsx): Example outputs obtained from the cGSA online platform.

Data file S7 (.xlsx): Raw experimental results related to Figures 2 and 4.